\documentclass[reprint,aip,prb]{revtex4-1}     
\usepackage{graphicx}

\begin{document}



\title{Development of a fast electromagnetic shutter for compressive sensing imaging in scanning transmission electron microscopy}

\author{A. B\'ech\'e}\email[Corresponding author: ]{armand.beche@uantwerpen.be}
\affiliation{EMAT, University of Antwerp, Groenenborgerlaan 171, 2020 Antwerp, Belgium}
\author{B. Goris}\affiliation{EMAT, University of Antwerp, Groenenborgerlaan 171, 2020 Antwerp, Belgium}
\author{B. Freitag}\affiliation{FEI Electron Optics, NL-5600 KA, Eindhoven, The Netherlands}
\author{J. Verbeeck}\affiliation{EMAT, University of Antwerp, Groenenborgerlaan 171, 2020 Antwerp, Belgium}

\begin{abstract}
The concept of compressive sensing was recently proposed to significantly reduce the electron dose
in scanning transmission electron microscopy (STEM) while still maintaining the main features in
the image. Here, an experimental setup based on an electromagnetic shutter placed in the condenser
plane of a STEM is proposed. The shutter blanks the beam following a random pattern while the
scanning coils are moving the beam in the usual scan pattern. Experimental images at both medium
scale and high resolution are acquired and then reconstructed based on a discrete cosine
algorithm. The obtained results confirm the predicted usefulness of compressive sensing in
experimental STEM even though some remaining artifacts need to be resolved.
\end{abstract}


\maketitle


\section{Introduction}\label{Part_intro}
One of the most challenging topics in modern transmission electron microscopy (TEM) is to perform
experiments on soft or beam sensitive materials as they suffer from irradiation damage that can
range from structural modification to the complete destruction of the sample. Such sample
modifications are even more problematic when 3D or/and analytical characterizations are involved.
In order to overcome this issue, a wide range of different approaches are being tested in the TEM
community like reduction of the kinetic energy of the fast electrons
\cite{rose_criteria_2007,muller_structure_2009,kaiser_transmission_2011}, improving the detection
efficiency of cameras
\cite{milazzo_initial_2011,bammes_direct_2012,li_electron_2013,mcmullan_comparison_2014}, time
resolved approaches
\cite{lagrange_single-shot_2006,aidelsburger_single-electron_2010,batson_unlocking_2011} and many
more. Relatively recently, the concept of compressive sensing was proposed to significantly reduce
the electron dose while maintaining all the important features in a TEM or Scanning TEM (STEM)
image \cite{binev_compressed_2012,stevens_potential_2014,li_wavelet_2014,saghi_reduced-dose_2015}.
Compressive sensing is based on the assumption that an image contains a significant amount of
redundancy and not all pixels in the image are independent. The condition for a proper image
reconstruction is that the original image has a sparse representation in a specific basis which
can be chosen prior to image reconstruction
\cite{candes_stable_2006,donoho_compressed_2006,candes_introduction_2008}. Therefore the image can
be approximated from a small subset of pixels randomly taken from the completely sampled image.
Depending on the redundancy, such reconstructed image can be very close to the fully sampled image
while considerably reducing the required dose. The process is somewhat comparable to image
compression algorithms that try to represent images with less information by exploiting the
redundancy present in a typical image.

Over the last few years, some studies have indeed proposed to apply compressive sensing to
transmission electron microscopy images using different types of reconstruction algorithms based
on Bayesan dictionary learning technique \cite{stevens_potential_2014}, wavelet frame based
\cite{li_wavelet_2014} or total variation inpainting \cite{saghi_reduced-dose_2015}. All this work
was done on virtually masked images starting from a fully sampled experimental or stored image and
applying a digital mask to it, taking out a number of often randomly selected pixels. This indeed
shows a great potential for compressive sensing but should be seen as an idealized simulation
assuming that experimental random pixel measurements would be possible. Implementing compressive
sensing in practice is complicated by the fact that typical scan engines in STEM microscopes are
not designed to be driven in a non-regular pattern as would be required. An alternative is to make
use of a beam shutter that can switch the electron beam on and off while keeping the conventional
regular scanning pattern. In this paper we present an experimental realization of such a beam
shutter based on electromagnetic deflection and demonstrate that we obtain experimental
compressive sensing in a STEM in both medium and high resolution. By shuttering the beam using a
pseudorandom generator, it was possible to acquire images with a limited number of pixels and
reconstruct them using the discrete cosine transform. This demonstrates that compressive sensing
became an experimentally viable technique in STEM opening up the predicted advantages of the
technique for experimental research.

\section{Experimental setup}\label{Part_exp_setup}

\begin{figure}[t]
\begin{center}
\includegraphics[width=\columnwidth]{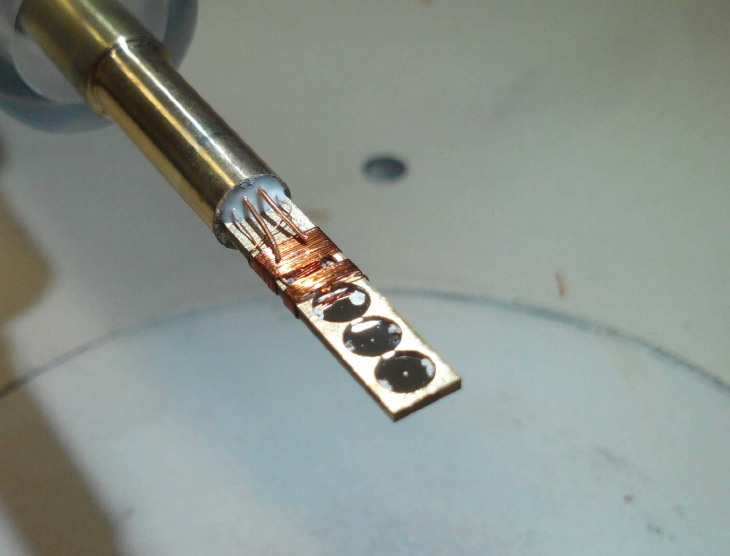}
\caption{\protect\textit{C2 electric contact aperture holder mounted with a
solenoid.}}\label{Aperture_holder}
\end{center}
\end{figure}

The aim in compressive sensing acquisition is to illuminate only parts of the sample. Scanning
imaging modes are especially well suited for this operation as images are acquired in a pixel by
pixel way. Mainly two strategies can be followed to achieve compressive sensing in such mode: (i)
blanking the beam in between two illuminated pixels or (ii) driving the scanning coils in a
specific way to jump from one pixel to a non-adjacent next pixel. Stevens et al.
\cite{stevens_potential_2014} achieved successful acquisitions with the second strategy in a SEM.
Doing this in a STEM is more complicated due to the higher beam location precision and scanning
speed that are typically required. Although less interesting in terms of total acquisition speed,
the blanking method has the distinctive advantage of a simpler hardware setup and will allow us to
test compressive sensing in STEM. In a TEM, the pre-specimen beam blanker is located at the gun
level but suffers from relatively slow response time, making it unattractive for the current
purpose of blanking the beam during scanning. Our first challenge was then to realize a
sufficiently fast beam shutter compatible with a typical microsecond range dwell times in STEM.
The most accessible locations in the illumination system of a modern TEM are the condenser
aperture holders. In the design of our TEM microscope, an FEI Titan3 equipped with both probe and
image Cs correctors and a Gatan Image Filter (GIF), the C1 apertures are located in the high
vacuum region of the gun making it improper for fast access and convenient operation. The C2
aperture further down the illuminating system was consequently the best place to place the
shutter. The first step was to design a completely new aperture holder with four electrical
feedthrough contacts. A picture of the custom built holder is displayed in Figure
\ref{Aperture_holder}. The four aperture slots are clearly visible together with four electrical
contacts. We deflect the electron beam making use of a simple solenoid wrapped around the first
condenser aperture. The solenoid produces a quasi-homogeneous magnetic field at the plane of the
aperture which deflects the electrons due to the Lorentz force. At the given winding density, it
turns out that a current of 250~mA is capable of deflecting a 300~keV focused probe about 350 nm
away from the sample area. A selected area (SA) aperture was introduced in the path of the beam
too prevent high angle diffraction signal from reaching the detector when the shutter blanks the
beam. The solenoid has a series resistance of $R_s = 0.75\Omega$ a self-inductance of $L_s =
0.3$~H and an estimated capacitance of 12~pF \cite{medhurst_hf_1947,medhurst_hf_1947-1}. These
parameters set the maximum switching speed which can be estimated as $\tau =  \sqrt(L_sC_s)$
leading maximum estimated switching speeds of 2~ns, much shorter than typical STEM dwell times.

\begin{figure}[b]
\begin{center}
\includegraphics[width=\columnwidth]{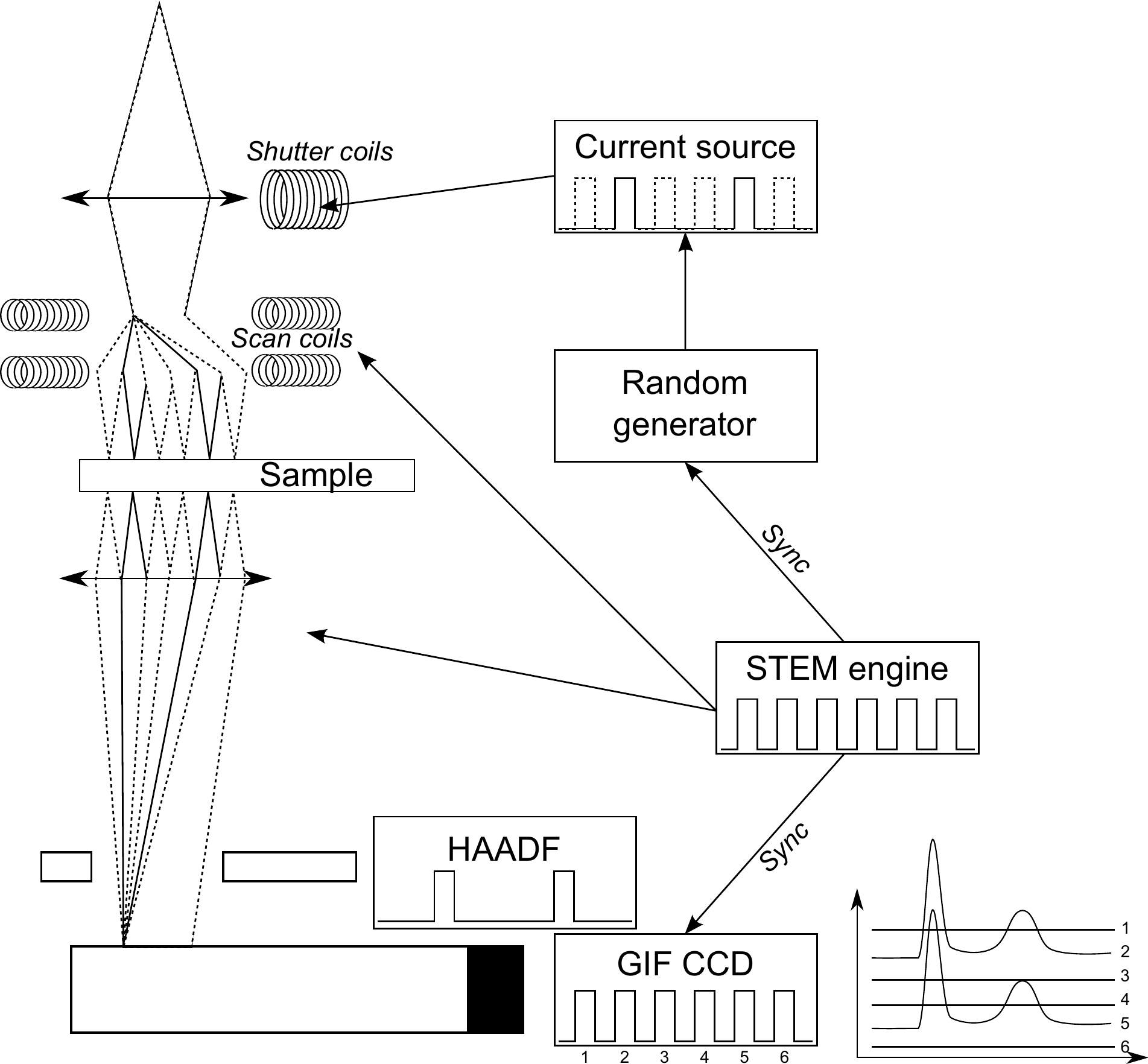}
\caption{\protect\textit{Schematic of the compressive sensing acquisition setup. The STEM engine
drives the STEM coils and synchronized the GIF CCD and the random generator. Depending on the
random number generator, the beam is either blanked (dashed trajectory) or not, leading to the
absence or presence of signal on the HAADF detector and zero loss peak in the EELS
map.}}\label{Experimental_setup}
\end{center}
\end{figure}

As the beam deflector was successfully implemented, we had to synchronize the STEM scan engine
with the beam deflection in order to properly shutter the beam at given pixel positions. An
Arduino \cite{_arduino_????} microcontroller unit linked to a switched current source was used to
drive the deflector coil synchronized with the shutter signal of the GIF CCD. The microcontroller
was then programmed to open or close the shutter based on a (pseudo) random
\cite{park_random_1988} generator in synchronization with the scan engine. A pixel will be
illuminated if the random generator with uniform probability distribution between 1 and 100 draws
a number higher than X, with X the average amount of unblanked pixels we want to obtain over the
whole scan. Figure \ref{Experimental_setup} displays a schematic of our experimental setup.

\begin{figure*}[t]
\begin{center}
\includegraphics[width=2\columnwidth]{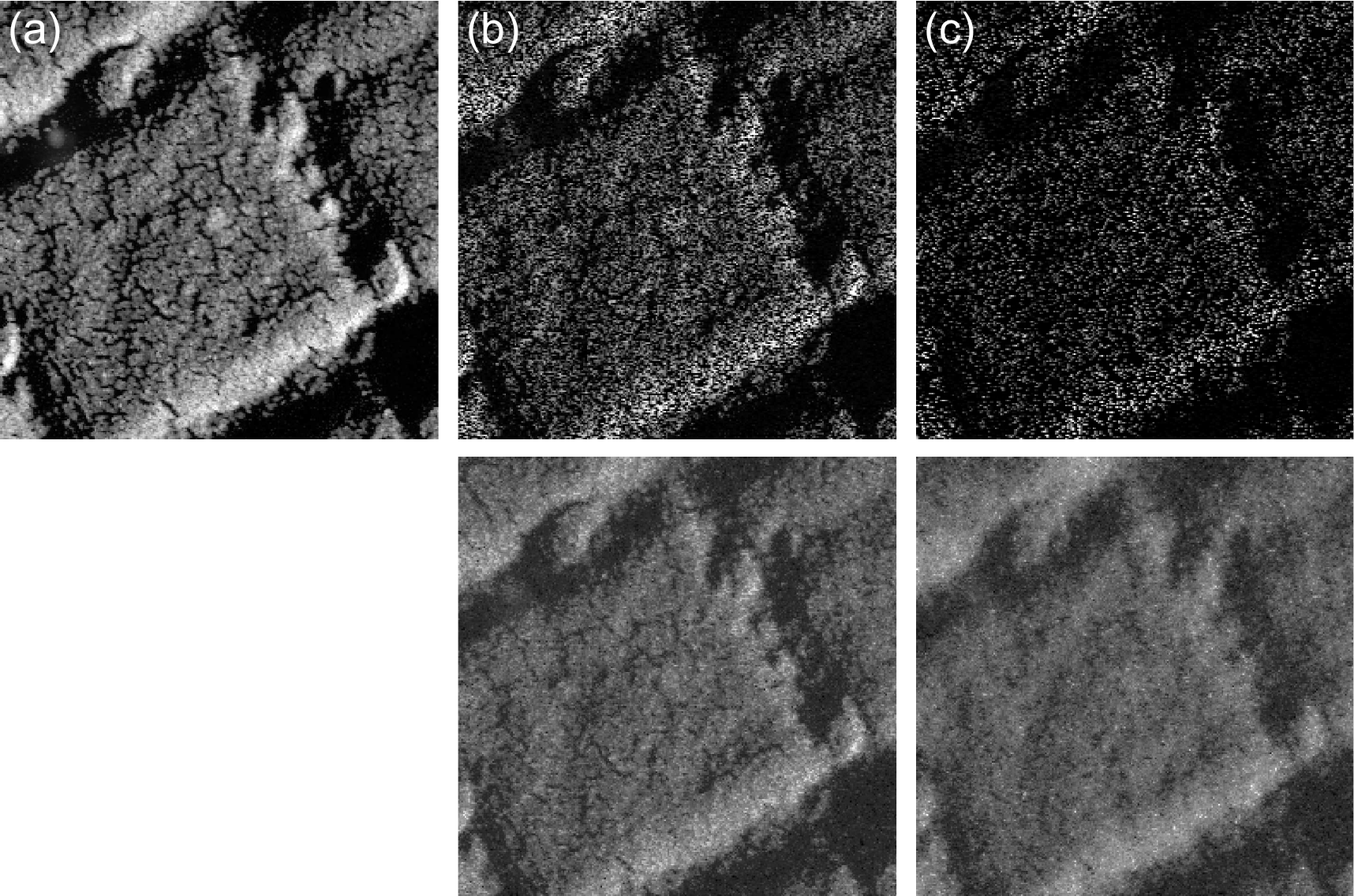}
\caption{\protect\textit{Experimental acquisition of STEM images at medium scale using (a) no beam
shuttering, (b) 50\% beam shuttering and (c) 80\% beam shuttering conditions. The reconstructed
images based on a discrete cosine algorithm are displayed below the experimental
images.}}\label{Cross_grating}
\end{center}
\end{figure*}

In order to correctly reconstruct the acquired sparsely sampled image, the reconstruction
algorithm needs to have access to the acquisition mask, namely knowing when the beam was blanked
or unblanked. As the storage space on the microcontroller was limited, we used a workaround to
obtain the shutter mask by acquiring a zero loss (ZL) EELS map together with the HAADF STEM image.
In addition to obtaining the applied shutter mask, this workaround allowed us to check how
efficiently the electron was shuttered by studying the intensity of the zero loss peak which
should ideally be high for unblanked pixels (the sample is electron transparent) and zero for
blanked pixels. In order to obtain a ZL peak of sufficient intensity, the single pixel exposure
time, for typical High Resolution STEM illumination settings, was set to 0.5 ms. Even though this
value is relatively high compared to typical dwell times used in STEM, it nevertheless allows us
to prove the setup works and to verify the efficiency of shuttering. In a later stage the
described workaround will disappear and the dwell time will be only limited by the maximum speed
at which the shutter can be driven which should be well in the microsecond to nanosecond range. In
order to reconstruct the images from the subsampled projections, an interpolation is required
filling the missing pixels in the images. This interpolation corresponds to solving the following
equations:

\begin{equation}
\hat{x} =arg min_x \parallel\Phi_x-b\parallel_{\ell2} with
\parallel\Psi_x\parallel_{\ell1}<\lambda
\end{equation}
Where $\hat{x}$ corresponds to the reconstructed image, $b$ equals the measured pixels and $\Phi$
is a subsampling operator that selects the imaged pixels. The operator $\Psi$ represents the
sparsifying transform that can be chosen prior to the reconstruction and $\lambda$ is a parameter
that can be adjusted according to the sparsity of the image after the sparsifying transform. In
this work, a discrete cosine transform is selected which is well suited for images showing a local
periodicity such as high resolution STEM projections. The reconstruction is implemented in Matlab
using the spgl1 algorithm \cite{van_den_berg_probing_2008,_spgl1:_????}. More elaborate transforms
can easily be implemented but we focused here on the experimental realization of the shutter.

\section{Results}\label{Results}

\begin{figure*}[t]
\begin{center}
\includegraphics[width=2\columnwidth]{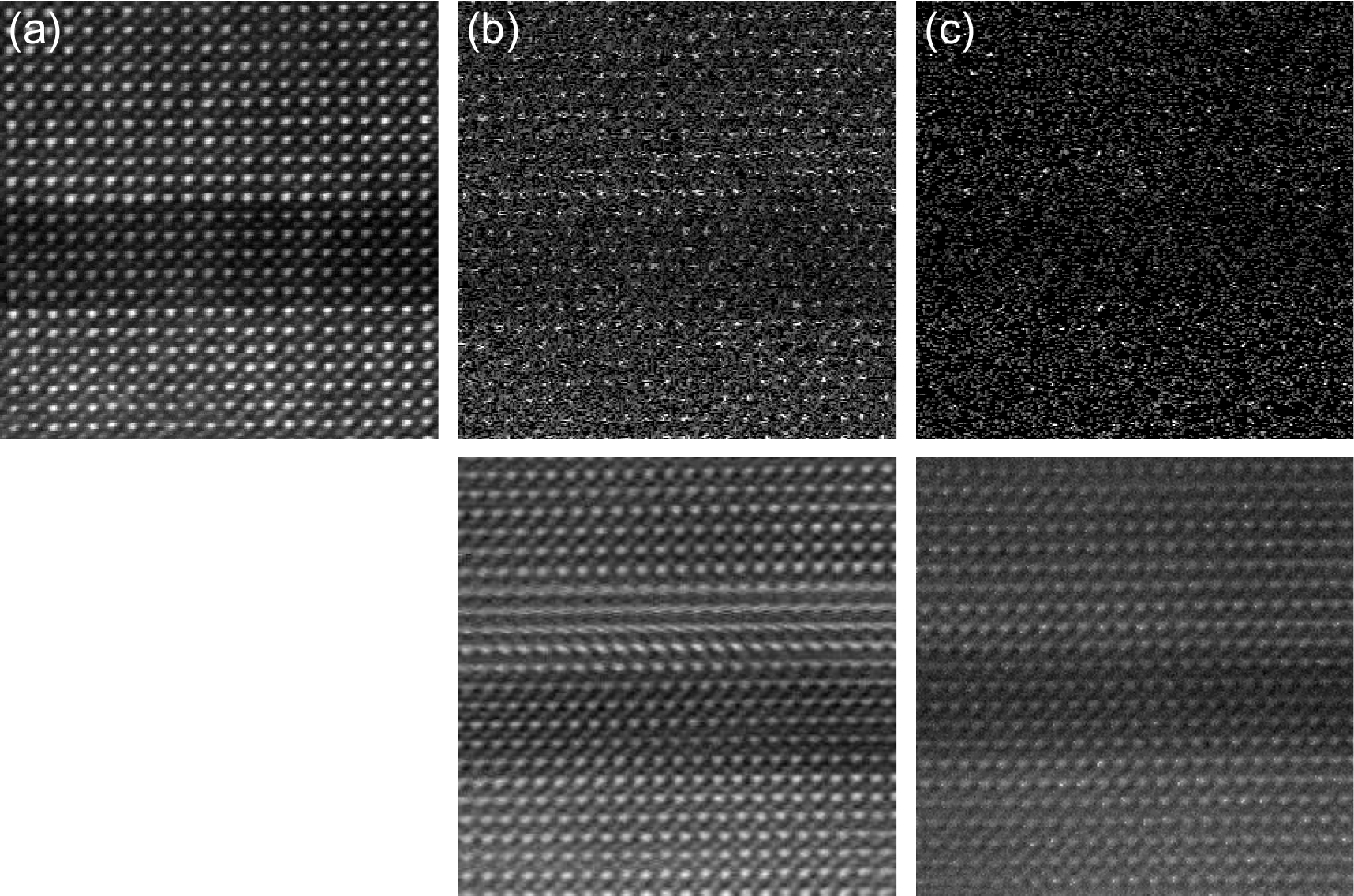}
\caption{\protect\textit{Experimental acquisition of STEM images at medium scale using (a) no beam
shuttering, (b) 50\% beam shuttering and (c) 80\% beam shuttering conditions. The reconstructed
images based on a discrete cosine algorithm are displayed below the experimental
images.}}\label{LSMO}
\end{center}
\end{figure*}

The effect of compressive sensing on STEM image acquisition and reconstruction was investigated
using two samples with rather different properties. On the one hand, medium resolution STEM
imaging was investigated on a standard gold cross grating sample. This sample has the advantage of
presenting a very high density of gold nanoparticles with quite complex agglomerated shapes. On
the other hand, HRSTEM was investigated on a complex perovskite oxide sample consisting of an
NbGaO$_3$ substrate covered with 6~atomic layers of SrTiO$_3$ and a 10~nm of LaSrMnO$_3$
\cite{liao_controlled_2015}. The lattice parameter is well above the theoretical resolution limit
of our instrument, thus insuring optimal conditions for the image reconstruction. Both samples
also have the advantage of being relatively beam hard allowing for the extra acquisition time
needed for the ZL spectrum mapping workaround. One could argue that both samples are rather far
from the beam sensitive samples one would expect when discussing compressive sensing, they allow
us however to study the feasibility of this new imaging technique without beam damage issues
complicating the interpretation of the results. Both samples were imaged with three different
acquisition schemes: no shuttering, 50\% shuttering and 80\% shuttering. The total dose is then
effectively reduced respectively by a factor of 2 and 5 in the different cases. The total frame
size was set to 256x256 pixels with a dwell time of 0.5~ms in standard HRSTEM illumination
conditions, being an acceleration voltage of 300~kV, a convergence angle of 20~mrad with a beam
current of 50~pA using a 20~$\mu$m C2 aperture. The simultaneous ZL EELS map was acquired with a
collection angle of 35~mrad using a dispersion of 0.25~eV/pixel and a 5~mm GIF entrance aperture.
The experimental images on the cross grating sample are regrouped in Figure \ref{Cross_grating}
together with their reconstructions based on the discrete cosine algorithm.

For the 50\% shuttered case, the main features of the image are retrieved in the reconstructed
image, even though the resolution decreased somewhat. The 80\% shuttered image reveals a lack of
detail and only the larger image features are reconstructed while the finer structural details are
lost. Note that in terms of redundancy, the cross grating is probably a very challenging case for
compressive sensing for the same reasons that this is an excellent sample to align an electron
microscope providing irregular features without favoring certain directions over others.

\begin{figure*}[t]
\begin{center}
\includegraphics[width=2\columnwidth]{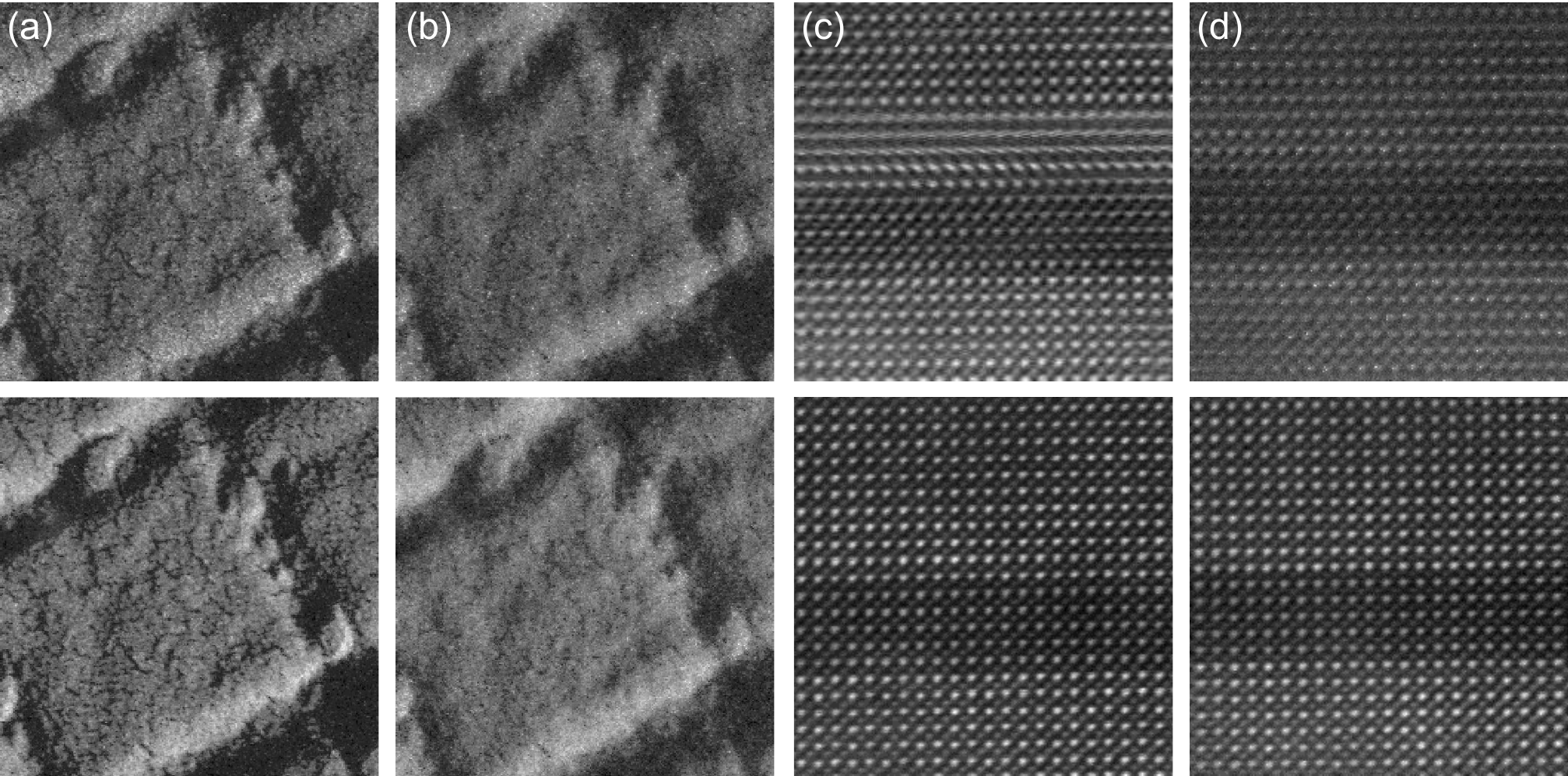}
\caption{\protect\textit{Comparison between reconstructed images from experiment (top row) and
from theory (bottom) row in four cases: (a) cross grating with 50\% shuttering time, (b) cross
grating with 80\% shuttering time, (c) HRSTEM with 50\% shuttering time and (d) HRSTEM with 80\%
shuttering time.}}\label{Discussion}
\end{center}
\end{figure*}

For the HRSTEM sample displayed in Figure \ref{LSMO}, both 50\% and 80\% shuttering cases are
still revealing acceptable high resolution information. The contrast of the lightest atoms tends
to significantly decrease in the 80\% shuttering case but all atoms remain visible. The presence
of the stripes in the middle of the 50\% shuttering case are due to some sample instability during
the acquisition and do not reflect any issues with the reconstruction algorithm. Because of the
longer acquisition times, the atomic lattice in Figures \ref{LSMO}b and \ref{LSMO}c is more
distorted by sample drift in comparison to Figure \ref{LSMO}a.

\section{Discussion}\label{Part_dicussion}
In order to discriminate shutter imperfections from reconstruction issues, we compare the
reconstructed images with the reconstructions obtained from a virtually shuttered image obtained
by applying a digital mask on the unshuttered experimental HAADF image as was typically done in
papers discussing compressive sensing so far
\cite{anderson_sparse_2013,li_wavelet_2014,saghi_reduced-dose_2015}. The results are shown in
Figure \ref{Discussion}, together with the reconstructed experimental images. In the case of the
cross grating sample, simulated and experimental images look very similar, with the small image
features being completely lost in the most sparsely sampled case. One can either incriminate the
reconstruction algorithm, which may fail to sufficiently exploit the sparsity in the observed
object, or it could be that the object itself simply doesn't have enough redundancy to be
accurately represented by sparse sampling. For the HRSTEM case, the difference between the
virtually shuttered images and the experimental ones is quite significant pointing towards shutter
artifacts. For the virtually shuttered images, in both the 50\% and 80\% shuttered cases, the
reconstructions are approaching the fully sampled image quite closely with the only noticeable
difference being a slight contrast reduction at sparser sampling. However, the experimentally
shuttered image reconstructions suffer from many more artifacts, going from a strong loss of
contrast to losing the light atoms all together. These artifacts can have many origins like
remaining synchronization and timing issues, sample drift caused by increased exposure times,
local sample charging issues affecting probe positioning and possibly others. As sample drift and
temporal instabilities of the instrument are related to the total acquisition time, they will
disappear when the setup is changed to exploit the full shutter speed as the workaround with the
ZL spectral acquisition is removed. Some of the mentioned artifacts could be overcome by using an
electrostatic shutter even though this implementation will likely have its own artifacts
\cite{ramasse_private_2015}.

\section{Conclusion}\label{Part_conclusion}
In this paper, we successfully demonstrate the implementation of compressive sensing in a TEM
making use of an electromagnetic beam shutter. Using a small solenoid placed in the condenser
system of the microscope, the beam can be independently shuttered for every pixel during a STEM
image acquisition. The reconstruction of the images from the experimentally obtained sparsely
sampled images shows that compressive sensing works significantly better on e.g. high resolution
images with much redundancy as compared to more irregular and less redundant images as
demonstrated with a cross grating sample. This is entirely expected but has to be kept in mind
when estimating the potential reduction in dose one could obtain from compressive sensing. At the
atomic scale, artifacts induced by sample drift significantly alter the result, but can be
entirely overcome in the future when lower dwell times are used. The speed of the present setup
remains insufficient for realistic use on beam sensitive samples but this limitation is imposed by
technological factors which can be overcome in the near future. If these remaining technological
challenges are overcome a reduction of dose of at least 5 times can be expected depending on the
type and sampling of the object. The implemented solution could offer a cost effective alternative
to e.g. a direct electron detection camera or can be combined with it in order to further reduce
the dose. Application in 3D tomographic acquisitions seem especially attractive as redundancy
between different projections could be exploited. This would be especially important in the case
of analytical 3D experiments.

\section{Acknowledgements}\label{Part_Acknowledgments}
A.B, B.G. and J.V. acknowledge funding from the European Research Council under the 7th Framework
Program (FP7), ERC Starting Grant No. 278510 VORTEX and No. 335078 COLOURATOM. A.B. and J.V.
acknowledge financial support from the European Union under the 7th Framework Program (FP7) under
a contract for an Integrated Infrastructure Initiative (Reference No. 312483 ESTEEM2). B.G.
acknowledges the Research Foundation Flanders (FWO Vlaanderen) for a postdoctoral research grant.
A.B., J.V. acknowledges funding from the GOA project SOLARPAINT and the POC project I13/009 from
the University of Antwerp. The Quantem microscope was partially funded by the Hercules Foundation.
We thank the Mesa+ laboratory at the University of Twente for the perovskite test sample.




\section{References}\label{Part_References}


\end{document}